# A single crystal $Co^{2+}$ fluoride pyrochlore antiferromagnet

J. W. Krizan[1*] and R. J. Cava[1]

[1]Department of Chemistry, Princeton University, Princeton, NJ 08544, USA

* Corresponding Author: jkrizan@princeton.edu

**Abstract**

We report the crystal growth, crystal structure and basic magnetic properties of a new cobalt-based pyrochlore, $NaSrCo_2F_7$. Na and Sr are completely disordered on the non-magnetic large atom *A* sites, while magnetic S=3/2 $Co^{2+}$ fully occupies the pyrochlore lattice *B* sites. This system exhibits an isotropic magnetic susceptibility with strong antiferromagnetic interactions ($\theta_{CW}$ = -127(1)), a large effective moment ($p_{eff}$ = 5.9(1) $\mu_B$/Co) and no spin freezing until 3 K. Thus $NaSrCo_2F_7$ is a geometrically frustrated antiferromagnet with a frustration index $f$ = -$\theta_{CW}$/$T_f$ ≈ 42. AC susceptibility, DC susceptibility, and heat capacity are utilized to characterize the spin freezing at low temperatures. We propose that $NaSrCo_2F_7$, compared here to the related material $NaCaCo_2F_7$, may be the realization of a frustrated pyrochlore antiferromagnet with weak bond disorder. The strong magnetic frustration at an easily accessible temperature and the availability of large single crystals make this one of the potentially interesting alternatives to the more commonly studied rare earth oxide pyrochlores.

Keywords: pyrochlore, frustrated magnetism, spin glass, fluoride

**Introduction**

The pyrochlore lattice is a classical geometry in the study of geometrically frustrated magnetism.[1] Many different types of magnetic ground states are found in these materials including long range ordered, spin ice, spin glass, and spin liquids. Some of the most commonly studied are the rare earth pyrochlores, of which many are oxides with the $A_2^{3+}B_2^{4+}O_7^{2-}$ formula where the magnetic rare earth (RE) sits on the *A*-site and the *B* site metal ion is non-magnetic. The $RE_2Ti_2O_7$ pyrochlores are the most studied of these due to the availability of large single crystals. Magnetic *B*-site pyrochlores should not be overlooked, however, as the *B*-site has the same magnetically frustrating pyrochlore geometry as the *A*-site. It is possible to place magnetic transition metal ions on the *B* site, but their charge has to be compensated by the appropriate adjustment of the chemical system. For divalent magnetic transition elements such as $Co^{2+}$, fluoride pyrochlores with a 1:1 mixture of large ions on the *A*-site are of interest. Here we describe the crystal growth and properties of the $NaSrCo_2F_7$ pyrochlore, which is a new addition to this family. We find that this strongly magnetically frustrated material can be grown as large single crystals by the Bridgeman-Stockbarger method in a floating zone furnace, which should facilitate future studies of its properties. Our characterization of the crystallography and magnetic properties show that Na and Sr are completely disordered over the *A*-site, and that this material exhibits a relatively large antiferromagnetic Curie Weiss temperature, a large magnetic moment, and spin glass behavior at approximately 3 K. We infer from the magnetic properties that the *A*-site disorder likely perturbs the Co-Co magnetic interactions, leading to weak bond disorder. Previous experimental work on $NaCaMg_2F_7$ shows that A-site disorder impacts the local B-site coordination[2] and theory suggests that this type of bond disorder can precipitate the spin glass ground state that is observed here. [3–5]

**Experimental**

Large single crystals of $NaSrCo_2F_7$ were synthesized and grown by a modified Bridgeman-Stockbarger method[6,7] in an optical floating zone furnace under a dynamic argon atmosphere (see supplemental material.) Pre-reaction of the elemental fluorides was done at 700°C in an anhydrous HF/argon atmosphere in a custom built, alloy 400 HF reactor. Air and water were rigorously excluded from the entire synthetic procedure. Extreme care should be, and was, exercised with HF gas as it is highly corrosive, toxic and incompatible with many materials. Structural characterization of a single crystal of $NaSrCo_2F_7$ was performed at 100 K (See supplemental material). Temperature dependent DC magnetization measurements were made in a Quantum Design Superconducting Quantum Interference Device (SQUID)-equipped Magnetic Property Measurement System (MPMS XL-5). Field-dependent magnetization, temperature dependent AC magnetization, and heat capacity measurements were completed in a Quantum Design Physical Property Measurement System (PPMS). All magnetization data were collected on the same crystal to eliminate the possibility of sample variation. Heat capacity measurements on $NaSrCo_2F_7$ were made by the heat relaxation method on a 12.8 mg single crystal mounted on a nonmagnetic sapphire stage with Apiezon-N grease. Low temperature heat capacity was collected with the Quantum Design $^3$He insert. The heat capacity of similarly prepared $NaCaZn_2F_7$[8] was subtracted to estimate the magnetic contribution.

**Results and Discussion**

The results of the single-crystal structure solution are given in tables I and II in the supplemental information. $NaSrCo_2F_7$ is a previously unreported, stoichiometric, fully fluorinated pyrochlore of the $A_2B_2F_7$ type where Na and Sr are completely disordered on the *A* site. No Na-Sr ordering could be detected (see reciprocal lattice in figure 1). Similarly, no *A-B* site mixing was detectable by single-crystal diffraction. The stoichiometry was similarly refined, and was found to be within error of the expected $NaSrCo_2F_7$. It was therefore fixed at that composition for the final refinement. As is found in many other pyrochlores, the *B*-site octahedron, here $CoF_6$, has six identical bond lengths and is slightly compressed along the <111> direction. This gives rise to two F-Co-F bond angles that differ from the ideal 90°. The Co-F coordination is shown in Figure 1. The deviation from the ideal angles is slightly smaller than is found for $NaCaCo_2F_7$[8,9] indicating the local $CoF_6$ crystal field is slightly more regular in the current case.

Powder-diffraction data of a ground, crystal fit very well with the structure determined from the single crystal diffraction data (Figure 1). Powder X-ray diffraction patterns taken over time (not shown) indicate a slow change of peak intensities with exposure to air over the course of a few days. This degradation of powder samples is likely due to atmospheric water vapor, mirroring the hygroscopic nature of the starting materials. As a precaution, single crystals were stored in a desiccator. However, no degradation of monolithic crystals was observed during the time scale of this research, likely due to their much smaller surface to volume.

The DC magnetization measurements show that $NaSrCo_2F_7$ is a highly frustrated Co-based pyrochlore with isotropic magnetization. The magnetization with an applied field of 2000 Oe oriented along the [100], [110], and [111] directions is shown in the inset of Figure 2. For the sake of clarity, the inverse susceptibility of the [100] direction and Curie-Weiss fit (χ=C/T-θ) from 150-300 K are shown Figure 2. The strength of the magnetic frustration is typically described in terms of the frustration index, $f = -\Theta_{CW}/T_c$, where T is the transition temperature and $\theta_{CW}$ is the Curie-Weiss temperature.[10] The Curie-Weiss temperature is an estimate of the net magnetic interaction strength. It is found to be -127(1) K for $NaSrCo_2F_7$. No magnetic transition is apparent until approximately 3 K, giving this material a frustration index *f* of ~42. This indicates $NaSrCo_2F_7$ is slightly less frustrated than $NaCaCo_2F_7$, where *f* = 56. Like $NaCaCo_2F_7$, this material has a very large effective moment for S=3/2 $Co^{2+}$ at 5.9 $\mu_B$, indicating a large orbital contribution to the magnetic moment. The magnetization vs. applied field at 2 K shows subtle curvature, but the magnetization is nowhere near saturation within the accessible range. The M vs. H behavior does not deviate significantly from a straight line near the 2000 Oe applied field employed in the characterization of the susceptibility.

Rearranging the Curie-Weiss Law to C/(χ|θ|)=T/|θ|-1 it is possible to construct a normalized, dimensionless plot that is scaled by the effective moment (C) and the strength of the magnetic interactions (|θ|).[11] This type of plot allows different materials to be easily compared, and Figure 3 compares $NaCaCo_2F_7$ and $NaSrCo_2F_7$ in this way. Ideal antiferromagnets would follow the line y=x+1 (shown as a dashed line), with indications of magnetic ordering on the order of T/θ ~ 1. That is clearly not the case here for either material. Antiferromagnetic or ferromagnetic correlations at lower temperatures in excess of expectations from simple Curie-Weiss behavior are manifested as positive and negative deviations from y=x+1, respectively. Figure 3 shows $NaCaCo_2F_7$ and $NaSrCo_2F_7$ have very similar properties, but $NaCaCo_2F_7$ maintains more nearly ideal Curie-Weiss behavior to lower normalized temperatures. Deviations begin to appear near 0.3 T/θ before the material exhibits increased ferromagnetic fluctuations as it approaches its spin freezing transition well below a T/θ of 0.1. On the other hand, $NaSrCo_2F_7$ appears to show small antiferromagnetic deviations from ideal behavior before

deviating strongly from ideal Curie-Weiss behavior with larger ferromagnetic fluctuations as it approaches its spin-glass transition. For both materials, the deviations from the ideal Curie-Weiss law at temperatures above the spin-freezing transitions indicate the presence of competing interactions within the magnetic sublattice.

Figure 4 further examines the low temperature magnetization of NaSrCo$_2$F$_7$ near the magnetic freezing at 3 K. The AC susceptibility shows the transition temperature shifts to a higher temperature with increasing frequency (10 and 50 Hz data omitted for clarity). No additional features were observed in the AC susceptibility in the temperature range of 4-300 K (data not shown). The transition temperature was taken as the maximum of χ'. The shift of the freezing temperature (ΔT$_f$ = 0.24 K) with frequency (ω) can be described by the expression $\frac{\Delta T_f}{T_f \Delta log\omega}$, which is used to characterize spin glasses, and spin glass like materials.[12] NaSrCo$_2$F$_7$ gives a value of 0.027, which is very close to that seen for NaCaCo$_2$F$_7$ (0.029) and falls within the range expected for an insulating spin glass.[8,12] The transition in the DC magnetization is shown in the middle of Figure 4. The transition in NaSrCo$_2$F$_7$ shows significant hysteresis at an applied field of 200 Oe along the [100] direction. Similar bifurcation is seen with the field applied along the [110] and [111] crystallographic directions (not shown). This is a further indication of a glassy transition.

To further parameterize the spin-glass behavior, the Volger-Fulcher law, $T_f = T_0 - \frac{E_a}{k_b}\frac{1}{\ln(\tau_0 f)}$, can be used to relate the freezing temperature and frequency to extract the intrinsic relaxation time (τ$_0$), the activation energy of the process (E$_a$), and "the ideal glass temperature" (T$_0$). Under the circumstances, with few data points and limited temperature resolution, insufficient data exist to obtain a meaningful fit to all variables. As such, the intrinsic relaxation time (τ$_0$) was set to a physically meaningful value of 1x10$^{12}$ s, which also allows a direct comparison to NaCaCo$_2$F$_7$. The resulting fit gives E$_a$ = 1.3(1)x10$^{-3}$ Ev and T$_0$ = 2.6(1) K, and is presented in the bottom plot of Figure 4. The "ideal glass temperature" can be interpreted as either relating to the cluster interaction strength in a spin glass, or relating to the critical temperature of the underlying phase transition that is dynamically manifesting at T$_f$[12]. The maximum and bifurcation in the DC susceptibility at 3 K is taken as an estimate of the freezing temperature and is in good agreement with the value of T$_0$ found here. This same value used to calculate the frustration index of 42, as mentioned above. Not surprisingly, the activation energy of the process (1.3(1)x10$^{-3}$ eV) is very close to that found for NaCaCo$_2$F$_7$ (1.0x10$^{-3}$ eV).

Further characterization of the magnetic spins is accomplished through analysis of the heat capacity. The top panel of Figure 5 shows the magnetic heat capacity for NaSrCo$_2$F$_7$ and NaCaCo$_2$F$_7$ as obtained by subtracting the nonmagnetic analog NaCaZn$_2$F$_7$[8]. For NaSrCo$_2$F$_7$, the heat capacity of the nonmagnetic analog was scaled by 2.5% as an estimate of the phonon contribution. This comparison of the two data sets shows the heat capacity of NaSrCo$_2$F$_7$ is greater between 10 and 60 K, approaching the spin freezing temperature. Below 10 K, there is a broad maximum in the magnetic heat capacity slightly above the freezing temperature seen in the magnetic measurements. This peak is similar in shape to that of NaCaCo$_2$F$_7$ and its position indicative of spin glass behavior.[10] The inset to the bottom panel of the figure shows a close up of the heat capacity below the magnetic transition.

It sometimes is possible to learn about the elementary excitations present in the system and to infer the ground state.[10] The expected behavior of a spin glass is C(T) ∝ T below the transition. In this particular case, curvature is evident at this scale, the scale shown in Figure 5, and also on T$^2$ or T$^{3/2}$ scaling. Below 1 K, there appears to be the beginning of another feature in the heat capacity. This could be a Schottky

Anomaly at lower temperatures, though further work is necessary to determine its origin. Regardless, the presence of this second feature complicates this analysis and prevents a conclusion about the spin excitations from the current data.

The integration of the magnetic heat capacity shows saturation by 70 K and is compared to the values expected for Ising and S = 3/2 Heisenberg systems, R ln(2) and R ln(2S+1) respectively. A significant amount of entropy is frozen out near the spin-glass transition, and the overall amount of entropy frozen out is close to that expected for a two-state magnetic system. The difference in heat capacity between 20 and 60 K, and consequently the additional entropy of $NaSrCo_2F_7$, could mirror the differences in magnetic susceptibility and be related to the small antiferromagnetic deviations of $NaSrCo_2F_7$ seen in Figure 3, but future work will be necessary to probe this in further detail.

**Conclusion**

Despite the strong antiferromagnetic magnetic interactions in the $NaSrCo_2F_7$ pyrochlore, no spin freezing is found until low temperatures, indicating that this is a highly frustrated magnetic system. No ordering is observed until the glassy transition at 3 K. Fitting the AC susceptibility data near the spin freezing temperature with the Volger-Fulcher law gives results commensurate with an insulating spin glass and in close proximity to the characteristics derived for the related material $NaCaCo_2F_7$. Heat capacity measurements corroborate the spin glass behavior and indicate that the system has a total entropy loss near R ln(2) due to the freezing of the spins. As compared to $NaCaCo_2F_7$, there is more entropy released in $NaSrCo_2F_7$ between 20 and 60 K. This extra entropy mirrors slight deviations from ideal Curie-Weiss behavior at the same temperatures and may be of interest for future studies.

The crystal structure determination indicates that Na and Sr are fully disordered on the A site of the $NaSrCo_2F_7$ pyrochlore, leading to a random nonmagnetic second neighbor environment around the magnetic $Co^{2+}$ cation. As was shown in a Raman study of the related fluoride pyrochlore, $NaCaMg_2F_7$, this disorder lowers the local symmetry of the B-site.[2] This is likely the case in $NaSrCo_2F_7$ as well, though a similar investigation would be needed to confirm this. Theoretical studies of the Heisenberg antiferromagnet on the pyrochlore lattice indicate that weak randomness in the exchange interactions (i.e. magnetic bond disorder) can precipitate a spin glass ground state, with the spin glass temperature set by the strength of the bond disorder.[3–5] As compared to the Na-Ca disorder seen in $NaCaCo_2F_7$, the Na-Sr disorder in the current material may be stronger given the greater contrast of ionic radii. This may explain why a small shift of the transition temperature from 2.4 K (Na-Ca) to 3.0 K (Na-Sr) is found. Future work is necessary to fully understand the differences between $NaCaCo_2F_7$ and $NaSrCo_2F_7$, but it is clear that the magnetic properties are robust. Both exhibit strong magnetic frustration, strong magnetic interactions, and very large moments for S=3/2 $Co^{2+}$. $NaSrCo_2F_7$ further illustrates the usefulness of transition metal fluoride pyrochlores for the study of geometrically frustrated magnetism. The availability of large, high quality single crystals of the fluoride pyrochlores should facilitate future studies.

**Acknowledgements**

This research was conducted under the auspices of the Institute for Quantum Matter at Johns Hopkins University, and supported by the U. S. Department of Energy, Division of Basic Energy Sciences, Grant DE-FG02-08ER46544.

**Figure captions:**

**Figure 1:** (Color on-line) The top left panel presents a single crystal slice displaying the color of NaSrCo$_2$F$_7$ next to a larger oriented crystal. The top center panel shows a precession image from single crystal diffraction of the (0kl) plane. This illustrates the absence of a super lattice that could indicate long range structural ordering beyond that of the ideal pyrochlore structure. The top right panel shows the slightly compressed CoF$_6$ octahedra with the two different F-Co-F bond angles indicated. The angles towards the <111> axis of compression are 96.1° and those orthogonal to this axis are 83.9°, with all Co-F bond distances equivalent at 2.05 Å. The bottom panel shows the representative powder diffraction pattern of a crushed NaSrCo$_2$F$_7$ crystal confirming that the bulk material is a pure A$_2$B$_2$F$_7$ pyrochlore. Expected reflections for the pyrochlore are marked with green tics.

**Figure 2:** (Color on-line) The DC magnetic susceptibility of NaSrCo$_2$F$_7$ is shown with an applied field of 2000 Oe parallel to the [100], [110], and [111] crystallographic directions. The main panel shows the inverse susceptibility of the [100] crystallographic direction, the Curie-Weiss fit, and extracted parameters. The lower right inset shows the raw DC susceptibility data indicating isotropic behavior. The upper left panel shows the [100] direction magnetization at 2 K as a function of field of applied field.

**Figure 3:** (Color on-line) A dimensionless, normalized plot of the magnetization, compares the behavior of NaSrCo$_2$F$_7$ to NaCaCo$_2$F$_7$ at temperatures near the curie Weiss θ and below. This is done by arranging the Curie-Weiss Law to C/(χ|θ|)=T/|θ|-1. The results of the Curie-Weiss fits are given in the main panel. NaSrCo$_2$F$_7$ deviates significantly more than NaCaCo$_2$F$_7$ from the ideal Curie-Weiss behavior (dashed line). The upper inset shows the full range of data. The lower inset shows the scaled susceptibility data at low temperatures.

**Figure 4:** (Color on-line) The top panel explores the frequency dependence of the freezing temperature in the *AC* susceptibility. The temperature dependent susceptibility with an applied field of 20 Oe and at different frequencies is shown. The middle panel shows bifurcation in the zero field cooled and field cooled DC susceptibility with an applied field of 200 Oe. The bottom panel shows the parameterization of the spin freezing (as determined by the *AC* susceptibility) and fit to the Volger-Fulcher law.

**Figure 5:** (Color on-line) The top panel shows the magnetic heat capacity of NaSrCo$_2$F$_7$ and NaCaCo$_2$F$_7$ obtained by the subtraction of the nonmagnetic analog, NaCaZn$_2$F$_7$. The inset shows the raw data comparison indicating significantly greater heat capacity of the Co-pyrochlores near the spin freezing transitions. The lower panel inset highlights the curvature in the heat capacity below the transition. The lower panel shows the integration of the magnetic heat capacity as compared to the Ising (R ln(2)) and Heisenberg (R ln(2S+1)) limits.

Figure 1:

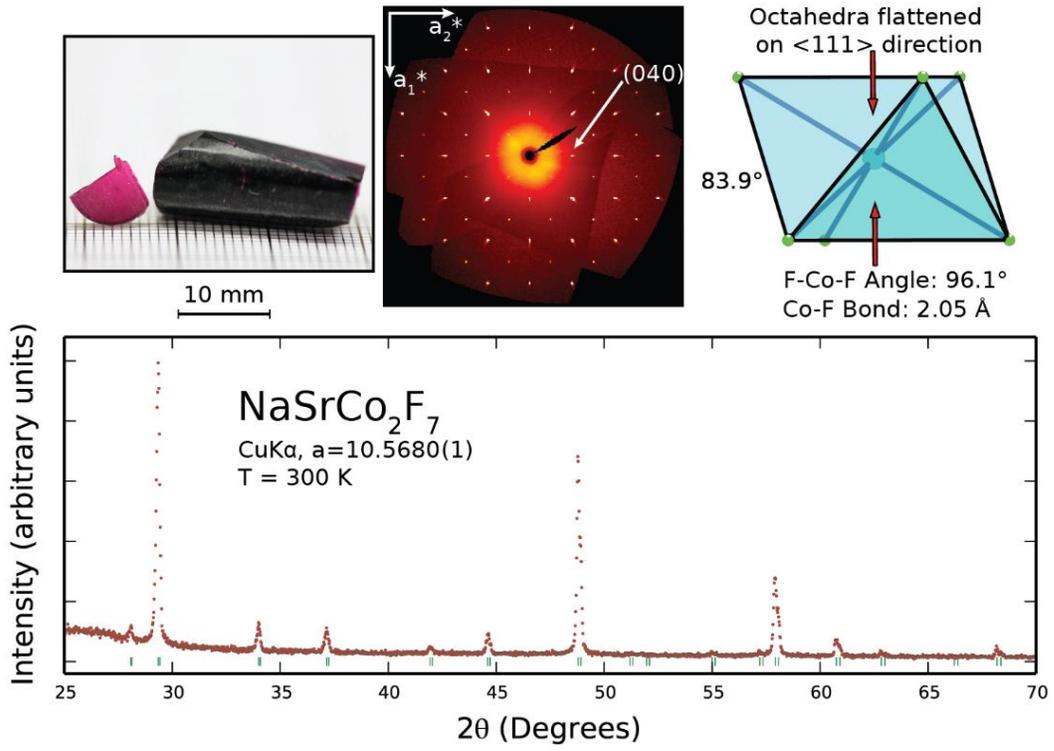

Figure 2:

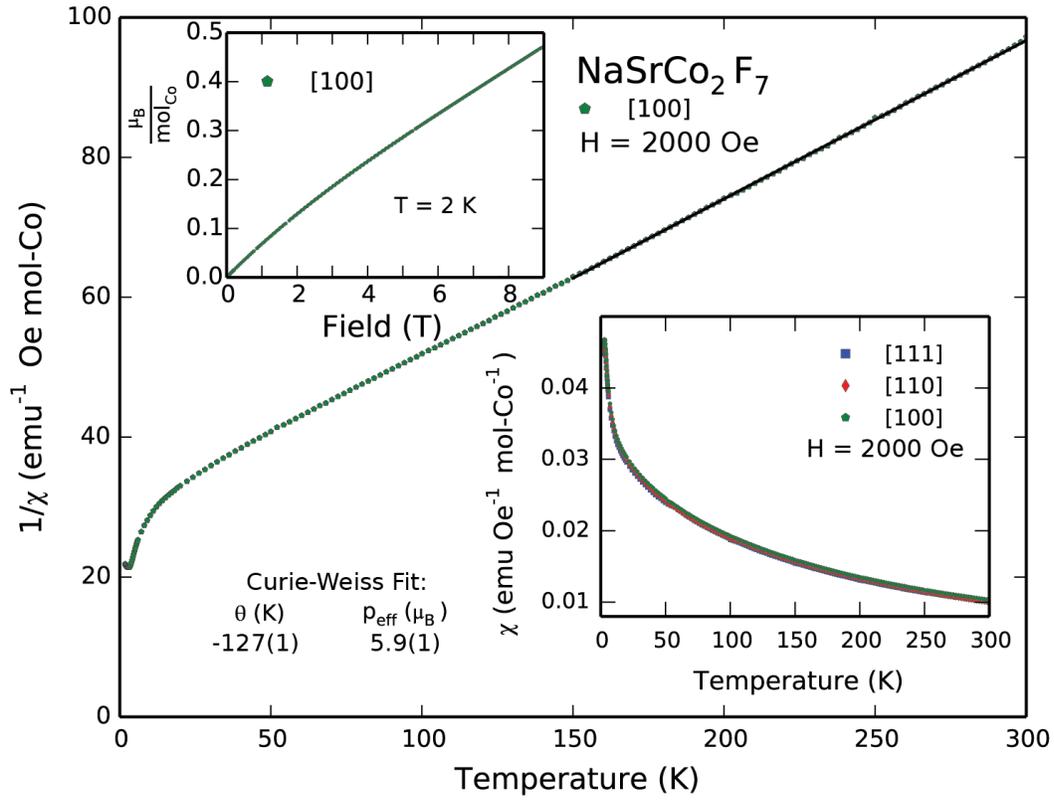

Figure 3:

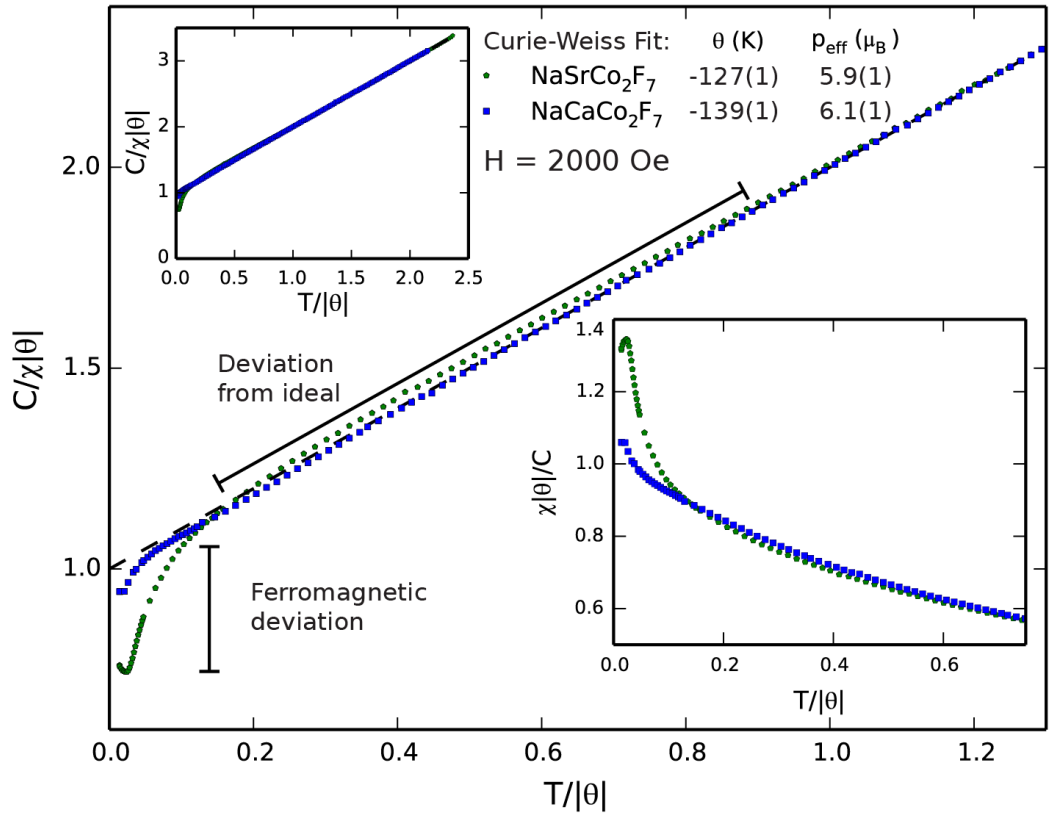

Figure 4:

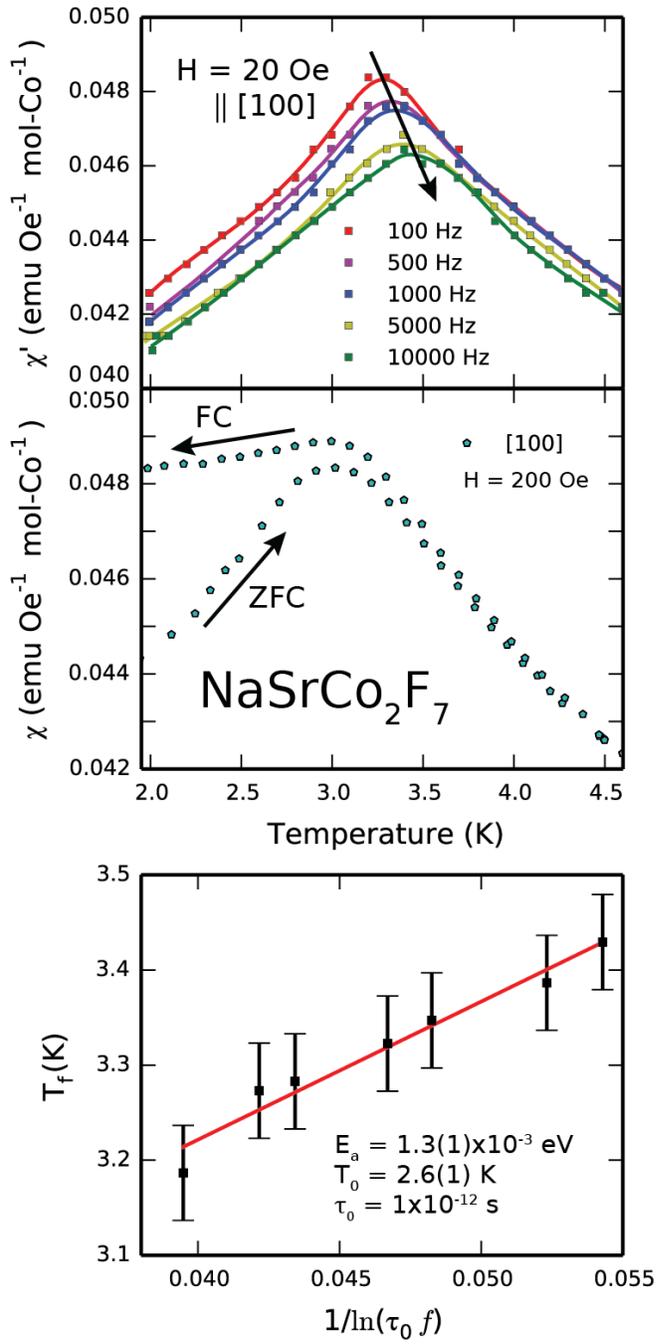

Figure 5:

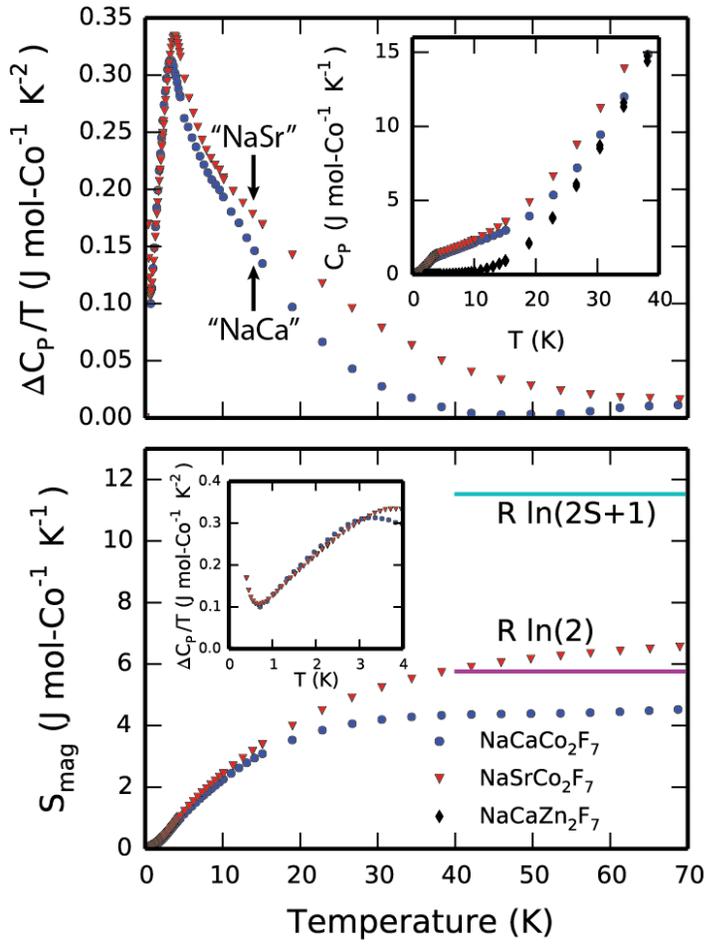

**Supplemental Material**

**Crystal Growth**

Extreme caution is required when working with HF. All work with these materials was conducted in a glove box in order to exclude air and water because of the hygroscopic nature of $CoF_2$. Prior to starting, NaF and $SrF_2$ were heated under dynamic vacuum at 175 °C for 24 hours. In a custom-built alloy 400 HF reactor, $CoF_2$ was dried under dynamic vacuum at 600 °C for eight hours and subsequently heated for 24 hours in an atmosphere of in-situ-generated anhydrous HF at 700 °C.

A stoichiometric mixture of the aforementioned fluorides was pre-reacted, again under an atmosphere of anhydrous HF, at 700 °C for 35 hours. The pre-reacted powder was loaded into a graphite mold, transferred to the optical floating zone and passed through the hot zone in order to synthesize and recrystallize the material. This approach is a modification of the Bridgeman-Stockbarger method. Crystal growth by the traditional floating-zone method is also possible under similar conditions. These methods benefit from the sharp temperature gradient possible in an image furnace.

The crystal growth was conducted in a Crystal Systems Corporation model FZ-T-10000-H-VI-VPO optical floating-zone furnace, equipped with four 500 W halogen bulbs under 8.5 bar of Ar pressure. The resulting temperature of the hot zone was approximately 950 °C. A translation rate of 4 mm/hr was successful. The large single crystal was oriented using a Laue camera (Multiwire Laboratories, model MWL110). Pieces several 2-5 mm in dimension were cut such that the magnetic field could be applied parallel to the [100], [110], and [111] crystal directions.

**Structure determination**

Single-crystal X-ray diffraction (SXRD) determination of the crystal structure of $NaSrCo_2F_7$ at 100 K was performed on a Bruker diffractometer equipped with an Apex II detector and graphite-monochromated Mo Kα radiation. Data collection was performed using the Bruker APEXII software package and subsequent reduction and cell refinement was handled with Bruker SAINT[13]. The crystal structure was determined using SHELXL-2013, as implemented through the WinGX software suite[14,15]. PXRD was collected at 300 K in a custom air-free sample holder, using a Rigaku Miniflex with diffracted beam graphite monochromator (Cu Kα), to confirm the bulk material in the boules of $NaSrCo_2F_7$ as having the pyrochlore structure. The PXRD pattern was fit with the Thompson-Cox-Hastings pseudo Voigt profile convoluted with axial divergence asymmetry through the FullProf software suite[16]. A unit cell of 10.5766 (7) Å was found for the bulk piece of the $NaSrCo_2F_7$ by least squares fitting of the PXRD pattern, in good agreement with the single crystal diffraction.

**Table 1: Single Crystal Data and Structural Refinement for NaSrCo$_2$F$_7$**

| | |
|---|---|
| Formula weight | 361.465 g/mol |
| Crystal System | Cubic |
| Space Group | $F\,d\,\bar{3}\,m$ (227, origin 2) |
| Unit Cell | a=10.545(4) Å |
| Volume | 1172.57(13) Å$^3$ |
| Z | 8 |
| Radiation | Mo Kα |
| T | 100 K |
| Absorption Coefficient | 14.801 |
| F(000) | 1328 |
| Reflections collected/unique | 3150/91 R$_{int}$ = 0.0231 |
| Data/Parameters | 91/11 |
| Goodness-of-fit | 1.176 |
| Final R indices [I>2σ(I)] | R$_1$=0.0213, wR$_2$=0.0584 |
| Largest diff. peak and hole | 0.36 and -0.57 e A$^{-3}$ |

**Table 2: Atomic positions and anisotropic thermal displacement parameters for NaSrCo$_2$F$_7$**

| | Site | x | y | z | Occ. | U11 | U22 | U33 | U23 | U13 | U12 |
|---|---|---|---|---|---|---|---|---|---|---|---|
| Na | 16d | 0.5 | 0.5 | 0.5 | 0.5 | 0.0175(5) | 0.0175(5) | 0.0175(5) | -0.0038(2) | -0.0038(2) | -0.0038(2) |
| Sr | 16d | 0.5 | 0.5 | 0.5 | 0.5 | 0.0175(5) | 0.0175(5) | 0.0175(5) | -0.0038(2) | -0.0038(2) | -0.0038(2) |
| Co | 16c | 0 | 0 | 0 | 1 | 0.0039(4) | 0.0039(4) | 0.0039(4) | -0.00032(15) | -0.00032(15) | -0.00032(15) |
| F(1) | 8b | 0.375 | 0.375 | 0.375 | 1 | 0.0190(14) | 0.0190(14) | 0.0190(14) | 0 | 0 | 0 |
| F(2) | 48f | 0.3285(2) | 0.125 | 0.125 | 1 | 0.0127(14) | 0.0091(8) | 0.0091(8) | 0.0037(9) | 0 | 0 |